\documentclass[letterpaper,aps,prl,twocolumn,showpacs,preprintnumbers,superscriptaddress]{revtex4}
\usepackage{graphicx}
\usepackage{epsfig}
\usepackage{bm}
\usepackage{amsmath}
\usepackage{amssymb}
\usepackage{latexsym}
\usepackage{color}

\begin{document}

\title{Crack Front Segmentation and Facet Coarsening in Mixed-Mode Fracture}
\author{Chih-Hung Chen}
\affiliation{Physics Department and Center for Interdisciplinary Research on Complex 
Systems, Northeastern University, Boston, Massachusetts 02115, USA}
\author{Tristan Cambonie}
\affiliation{Laboratoire FAST, Univ Paris Sud, CNRS, Universit\'e Paris-Saclay, F-91405, Orsay, France}
\author{Veronique Lazarus}
\affiliation{Laboratoire FAST, Univ Paris Sud, CNRS, Universit\'e Paris-Saclay, F-91405, Orsay, France}
\author{Matteo Nicoli}
\affiliation{Physics Department and Center for Interdisciplinary Research on Complex 
Systems, Northeastern University, Boston, Massachusetts 02115, USA}
\author{Antonio Pons}
\affiliation{Department of Physics and Nuclear Engineering, Polytechnic University of Catalonia, Terrassa,
Barcelona 08222, Spain}
\author{Alain Karma}
\affiliation{Physics Department and Center for Interdisciplinary Research on Complex 
Systems, Northeastern University, Boston, Massachusetts 02115, USA}

\date{\today}

\begin{abstract}
A planar crack generically segments into an array of ``daughter cracks'' shaped as tilted facets when loaded with both a tensile stress normal to the crack plane (mode I) and a shear stress parallel to the crack front (mode III). We investigate facet propagation and coarsening using in-situ microscopy observations of fracture surfaces at different stages of quasi-static mixed-mode crack propagation and phase-field simulations. The results demonstrate that the bifurcation from propagating planar to segmented crack front is strongly subcritical, reconciling previous theoretical predictions of linear stability analysis with experimental observations.  They further show that facet coarsening is a self-similar process driven by a spatial period-doubling instability of facet arrays with a growth rate dependent on mode mixity. 
Those results have important implications for understanding the failure of a wide range of materials. 
\end{abstract}


\maketitle

Crack propagation is a main mode of materials failure. Understanding and controlling this complex phenomenon continues to pose both fundamental and practical challenges. While quasi-static planar crack growth with a tensile stress normal to the fracture plane (mode I) is well-understood, geometrically much more intricate crack patterns can form in varied conditions  \cite{Bouchetal2010}. A few examples include 
thermal or drying stresses that can cause cracks to oscillate and branch \cite{YuseSano1993, HesPerRon95}, or re-organize into complex three-dimensional patterns \cite{GauLazPau10, MBGL13, Bourdinetal2014}, nonlinear elastic effects that can induce crack front instabilities even in mode I \cite{Baumbergeretal2008}, or the superposition of mode I and a shear stress parallel to the crack front (mode III). This mixed-mode I+III fracture is observed in a wide range of engineering and geological materials to produce arrays of daughter cracks, which are shaped as tilted facets and form by a geometrically complex crack front segmentation process 
\cite{S69,K70,PK75,HP79,PSD82,ST87,PA88,YM89,CP96,L97,LLM01b,LBFW08,LMR10,GO12,PR14,RCB14}. 
 
Recent theoretical progress has been made to characterize the crack-front instability leading to segmentation \cite{PK10,LKL11} and to describe the propagation of daughter-crack arrays \cite{LLK15}. However, theory and experiments have not produced a consistent picture. Griffith's energetic criterion \cite{G21} predicts that planar crack growth is possible when the elastic energy release rate
\begin{equation}
G=\frac{1}{2\mu} \left((1-\nu)K_I^2+K_{III}^2\right), \label{Gdef}
\end{equation} 
exceeds a critical material-dependent threshold $G_c$, where $K_I$ and $K_{III}$ are the mode I and mode III stress intensity factors (SIF), respectively, which characterize stress divergences near the crack front,  $\mu$ is the shear modulus and $\nu$ is Poisson's ratio. Phase-field simulations of brittle mixed-mode I+III fracture have revealed that planar growth is linearly unstable against helical deformations of the crack front, which couple in-plane and out-of-plane perturbations and develop nonlinearly into facets \cite{PK10}. A subsequent linear stability analysis in the framework of linear elastic fracture mechanics (LEFM) \cite{LKL11} has predicted that this helical instability should occur when $K_{III}/K_I$ exceeds a threshold
 \begin{equation}
  \left(\frac{K_{III}}{K_I}\right)_c = \sqrt{ \frac{ (1-\nu)(2-3\nu) }{ 3(2-\nu)-4\sqrt{2}\,(1-2\nu) } }, \label{K3K1cr}
\end{equation}
which only depends on Poisson's ratio. However, crack front segmentation is experimentally observed for $K_{III}/K_I$ values much smaller than this threshold  \cite{S69,RCB14}, or even vanishingly small \cite{PR14}.  
This apparent disagreement between linear stability analysis and experiment raises the question of whether LEFM and phase-field modeling are adequate theories to describe crack propagation in mixed-mode I+III brittle fracture. Also poorly understood is ``facet coarsening'', the progressive increase of facet width and spacing with propagation length from the parent crack.  Phase-field modeling \cite{PK10} and experiments \cite{CL2014} suggest that coarsening may be a self-similar process, but its precise mechanism and dependence on mode mixity are not well understood. 

\begin{figure}[t!]
\begin{center}
\includegraphics[width=0.47\textwidth]{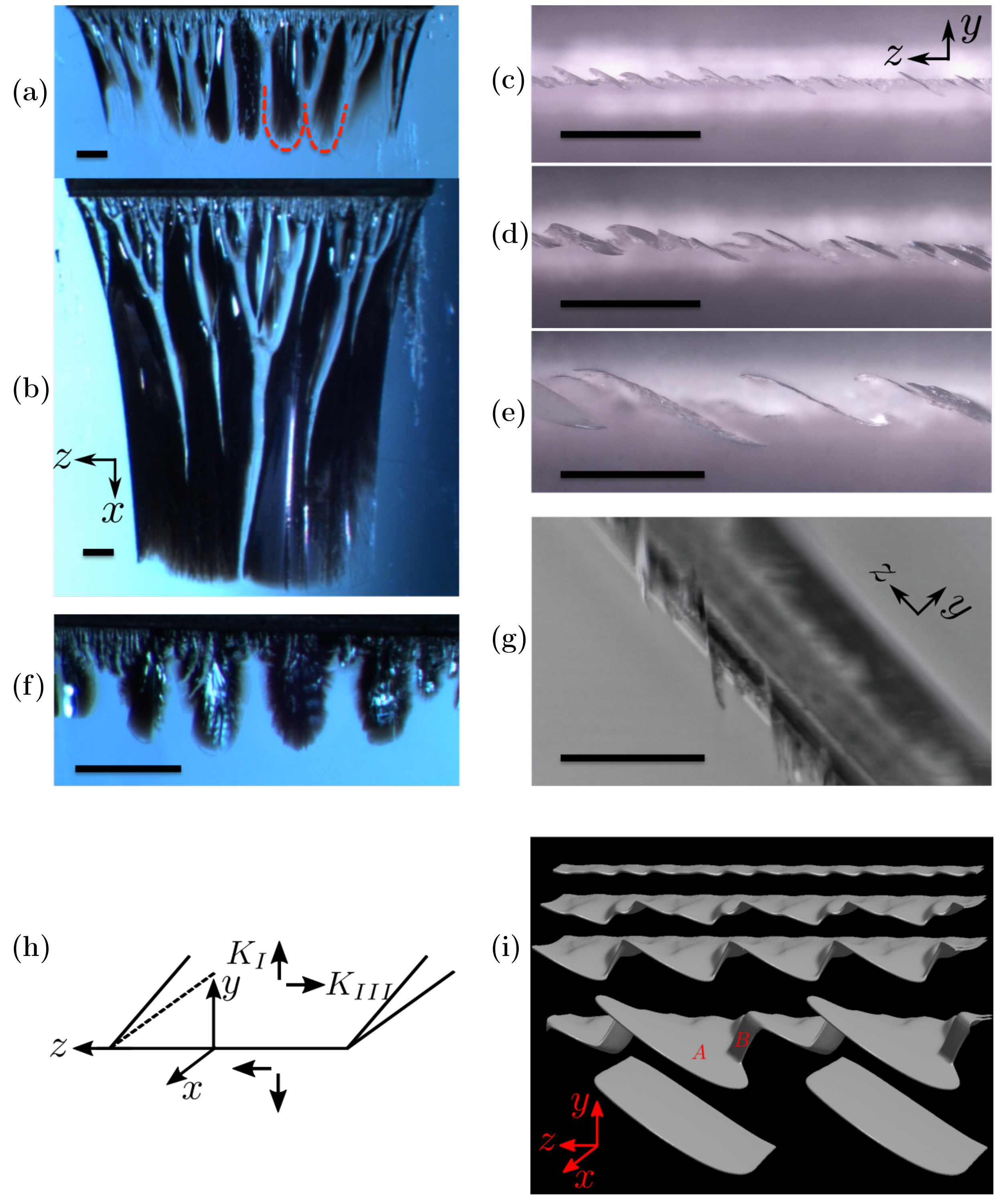}  
 \caption{(Color online). In-situ microscope images (a)-(g) of fatigue cracks in plexiglass at different stages of crack advance in mixed mode I+III loading depicted schematically in (h) and corresponding example of crack-front segmentation in phase-field simulation (i). $K_{III}/K_I\approx 0.3$ in (a)-(e) and $\approx 0.5$ in (f)-(g); (a), (b) and (f) are experimental views from a direction approximately perpendicular to the plane of the parent crack with facets propagating downwards, while views (c), (d), (e) and (g) are views with the crack propagation direction out of the page. Views (c), (d) and (e) correspond to different stages of crack advance increasing from (c) to (e). Broken (pristine) regions of the samples appear in black (light blue) or darker (lighter) grey depending on the viewing direction. The bar scale is 1 mm in all images. The red dashed lines in (a) highlight the curved fronts of two facets as guide to the eye; curved tips are clearly visible in (f). (i) Snapshots of phase-field fracture surfaces ($\phi=1/2$ surfaces) at different stages of crack advance increasing from top to bottom, showing that energetically favored A facets \cite{LLM01b} propagate ahead of B facets eventually outgrowing them completely.  Simulation parameters are $G/G_c=1.5$, $K_{III}/K_I=0.5$, and box dimensions $D_x=307\xi$, $D_y=100\xi$ and $D_z=200\xi$.}
 \vskip -0.8cm
\label{Fig1}
\end{center}
\end{figure}

In this letter, we investigate both facet propagation and coarsening by mixed-mode I+III fracture experiments that allow us to visualize in-situ complex crack morphologies during quasi-static propagation, thereby providing  much more detailed geometrical information on crack front evolution than conventional post-mortem fractography. Moreover, we carry out phase-field simulations of those experiments that allow us to relate experimental observations to LEFM theory. The results help resolve the puzzling discrepancy between linear stability analysis and experiments with regards to facet formation and shed new light on the coarsening process.

Experiments are carried out using plexiglas beams and a traditional three or four point bending setup \cite{Suppinfo}.
To introduce some amount of mode III, the initial planar notch in the sample is tilted at an angle from the mode I central plane of symmetry 
\cite{BucCheRic04, LBFW08}. A special procedure is used to initiate a sharp crack with a straight front  \cite{Suppinfo}. 
The corresponding values of the SIF for each angle and hence $K_{III}/K_I$ have been obtained by finite element calculations, which show
that $K_{III}/K_I$ varies between approximately 0.1 and 0.5 when the notch angle varies between $15^\circ$ and $45^\circ$, where zero angle corresponds to pure mode I loading. This range was selected because it contains the linear instability threshold $ (K_{III}/K_I)_c\approx 0.39$ predicted
by Eq. (\ref{K3K1cr}) for Poisson's ratio of plexiglass $\nu\approx 0.38$. 
Finite element calculations also show \cite{LBFW08} that  $K_{III}/K_I$
is reasonably constant away from sample edges, thereby allowing us to investigate crack propagation at constant $K_{III}/K_I$ along a wide section of the parent crack inside the sample.  
Several beams were broken by fatigue in the bending set-up \cite{Suppinfo}. 
The advantage of this cyclic type of loading is that the crack advance (i) is quasi-static, 
while leaving the crack path unchanged in comparison to the one obtained under monotonical increasing loading \cite{Lin93}
and (ii) controlled by the number of cycles so that complex crack morphologies can be observed in-situ at different stages of crack growth. 
Observations were made using a Leica binocular or a Keyence numerical microscope by transparency.

Examples of experimental images are shown in Fig. 1(a)-(g) for $K_{III}/K_I$ values of 0.3 and 0.5 corresponding to initial notch angles of $30^\circ$ and $45^\circ$, respectively. Those images reveal several important features. Firstly, facets have a finger-shape with curved tips and flat sides that is consistent with the shape predicted by phase-field simulations (Fig. 1(i) and Movie 1 of \cite{Suppinfo}). Secondly, facets form for values of $K_{III}/K_I$ both below and above the linear stability threshold $ (K_{III}/K_I)_c\approx 0.39$. Within optical resolution, only energetically favored type A facets are observed to emerge from the parent crack with a well-defined tilt angle $\theta$ from the original fracture plane.
Thirdly, facets coarsen by elimination of other facets leading to an increase of both facet width and facet spacing along the array with increasing propagation length. Coarsening is clearly visible from top views in Fig. 1(b) and in the sequence Fig. 1(c)-(e), which moreover shows that surviving facets maintain the same angle while overgrowing others. Additional views are given in \cite{Suppinfo}.

\begin{figure}[t!]
\begin{center}
\includegraphics[width=0.47\textwidth]{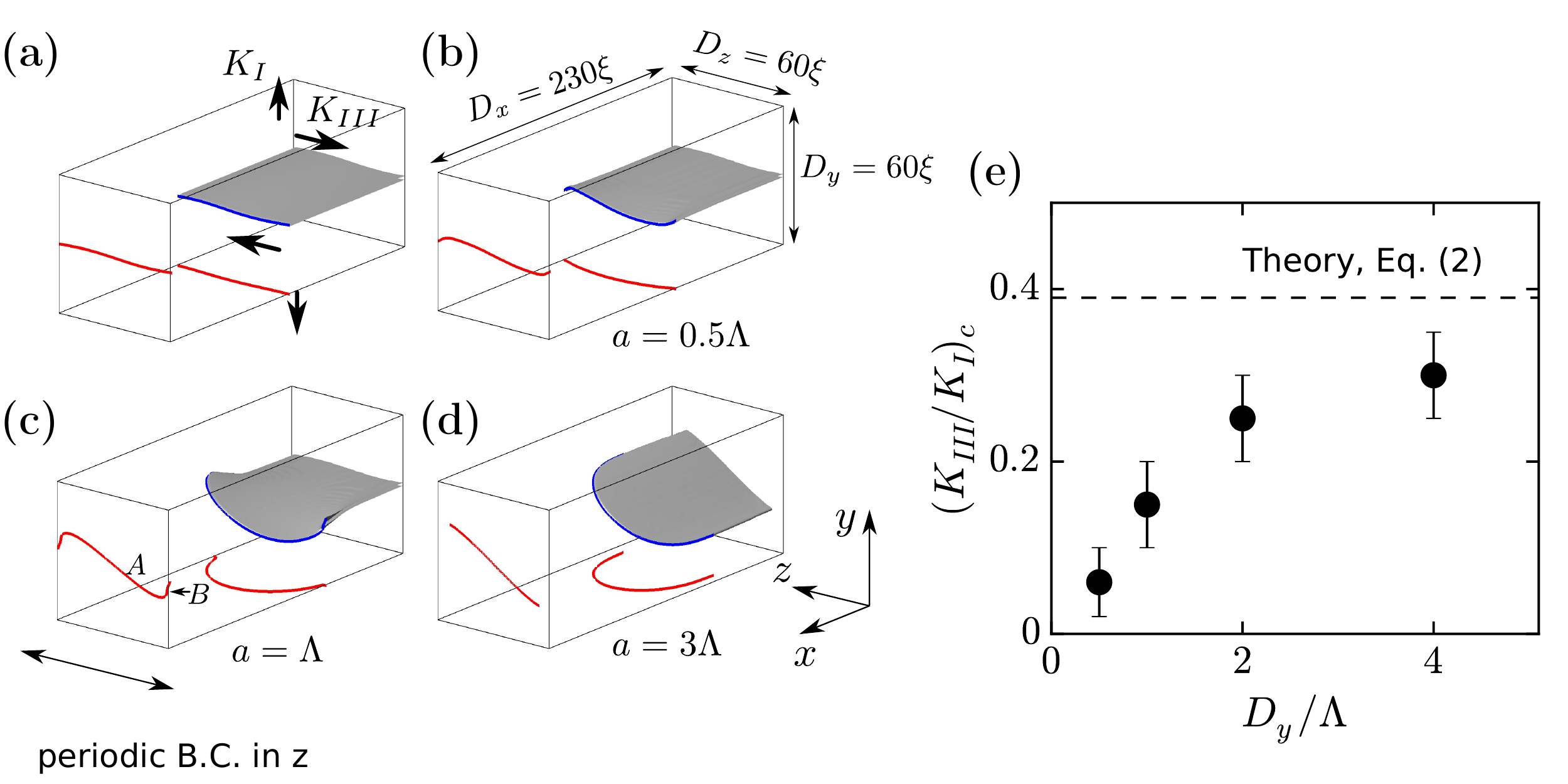} 
\caption{ 
(Color online). Snapshots phase-field simulations illustrating the destabilization of planar crack growth for $K_{III}/K_{I}=0.4$. The crack propagation length $a$ increases from (a) to (d) and both the crack front (blue lines) and its in-plane and out-of-plane projections (red lines) are shown. (e) Plot of linear instability threshold $\left(K_{III}/K_{I}\right)_c$ versus $D_y/\Lambda$. Planar growth is unstable (stable) above (below) the filled circles, where error bars reflect the uncertainty in stability threshold resulting from the fact that $K_{III}/K_I$ was increased in finite steps in the simulations. $K_{III}/K_I$ values corresponding to the top (bottom) of each error bar were simulated and found to yield unstable (stable) propagation. In all simulations, $G=1.5G_c$, $D_x=230\xi$ and $D_z=\Lambda=60\xi$.}
\vskip -0.8cm
\label{Fig2}
\end{center}
\end{figure}

Simulations were carried out using a phase-field model of brittle fracture that, like gradient damage models \cite{BFM00,Bourdinetal2014},  regularizes stress-field divergences on a process zone scale $\sim \xi$ around the crack front. All energy dissipation takes place on a characteristic timescale $\tau$ \cite{KKL01}.  
As shown by an asymptotic analysis of the phase-field model in the limit where $\xi$ is much smaller than all other dimensions \cite{HK09}, fracture in this model is governed by standard crack propagation laws assumed in the LEFM theoretical framework, namely Griffith's criterion and vanishing mode II SIF \cite{GS74}. 
Since we are primarily interested in modeling crack evolution in a region away from the experimental sample boundaries where $K_{III}/K_I$ is approximately uniform \cite{LBFW08,CL2014}, we carried out simulations in a rectangular slab geometry of length $D_x$,  width $D_y$ and height $D_z$, defined in Fig. \ref{Fig2}(b), with the origin defined at the center of the slab. We impose fixed displacements at $y=\pm D_y/2$, $u_y(x,\pm D_y/2,z)=\pm \Delta_y$ (mode I) and $u_z(x,\pm D_y/2,z)=\pm \Delta_z$ (mode III),  periodic boundary conditions in $z$ that allow us to model a periodic array of daughter cracks infinite in $z$ \cite{PK10}. We use a ``treadmill'' that adds a strained $(y,z)$ layer at $x=D_x/2$ and removes a layer at $x=-D_x/2$ when the crack has advanced by one lattice spacing. This allows us to simulate crack propagation lengths much longer than $D_x$ ($a\gg D_x$), thereby modeling propagation in a slab infinitely long in $x$ \cite{Suppinfo}. We also choose $D_x\ge 2.5 D_y$ to eliminate the influence of the two end-boundaries of the slab ($x=\pm D_x/2$) on the central region of the slab ($|x| \ll D_x$) where the average crack front position is maintained by the treadmill. 
Standard expressions of linear elasticity are used to relate $\Delta_y$ and $\Delta_z$ to the SIF \cite{Suppinfo} and therefore to $K_{III}/K_I$ and $G/G_c$ where $G_c\approx 2\gamma$ (twice the surface energy) is known in the phase-field model \cite{KKL01,HK09}. All simulations are performed with $\nu=0.38$ of plexiglass. We simulated both quasi-static propagation, where the elastic field is relaxed at each time step of crack advance, and dynamic propagation by solving the full elastodynamic equations. Both sets of simulations yielded similar results for the range $G/G_c\le 1.5$ where the ratio of the crack propagation speed to the shear wave speed $v/c\le 0.3$ is small enough to neglect inertial effects \cite{Suppinfo}.

We first carried out simulations to check quantitatively the theoretical prediction of Eq. (\ref{K3K1cr}). For this purpose, we slightly perturbed the planar parent crack with a small amplitude helical perturbation of the form $\delta x_{\rm{front }}+i\delta y_{\rm{front }}=A_{0}e^{-ikz}$, where $\delta x_{\rm{front }}$ and $\delta y_{\rm{front }}$ indicate the $x$ and $y$ components of deviations of the front from the reference planar crack, respectively, and $k=2\pi/D_z$ fits one wavelength $D_z=\Lambda$ of the perturbation in the periodic domain in $z$. The stability of planar crack propagation is then determined by tracking the amplitude of the perturbation that grows or decays exponentially in time \cite{Suppinfo} if propagation is unstable, as illustrated in Fig. 2(a)-(d), or stable, respectively. Simulations were carried out by increasing $K_{III}/K_I$ in small steps to determine the threshold $(K_{III}/K_I)_c$, and repeating this procedure for increasing values of $D_y/\Lambda$  
to quantify finite size effects. Fig. 2(e) shows that $(K_{III}/K_I)_c$ increases monotonously with $D_y/\Lambda$ and approaches a value reasonably close to the prediction $(K_{III}/K_I)_c\approx 0.39$ of Eq. (\ref{K3K1cr}) in the large system size ($D_y/\Lambda\gg 1$) limit. Consistent with the result of Fig. 2(e), an examination of strain fields shows that finite size effects becomes negligible when $D_y/\Lambda \ge 2$ \cite{Suppinfo}. We conclude that LEFM theory (Eq. (\ref{K3K1cr})) and phase-field modeling predict similar linear instability thresholds in the large system size limit, and that facets are experimentally observed well below this threshold.

\begin{figure}[t!]
\begin{center}
\includegraphics[width=0.47\textwidth]{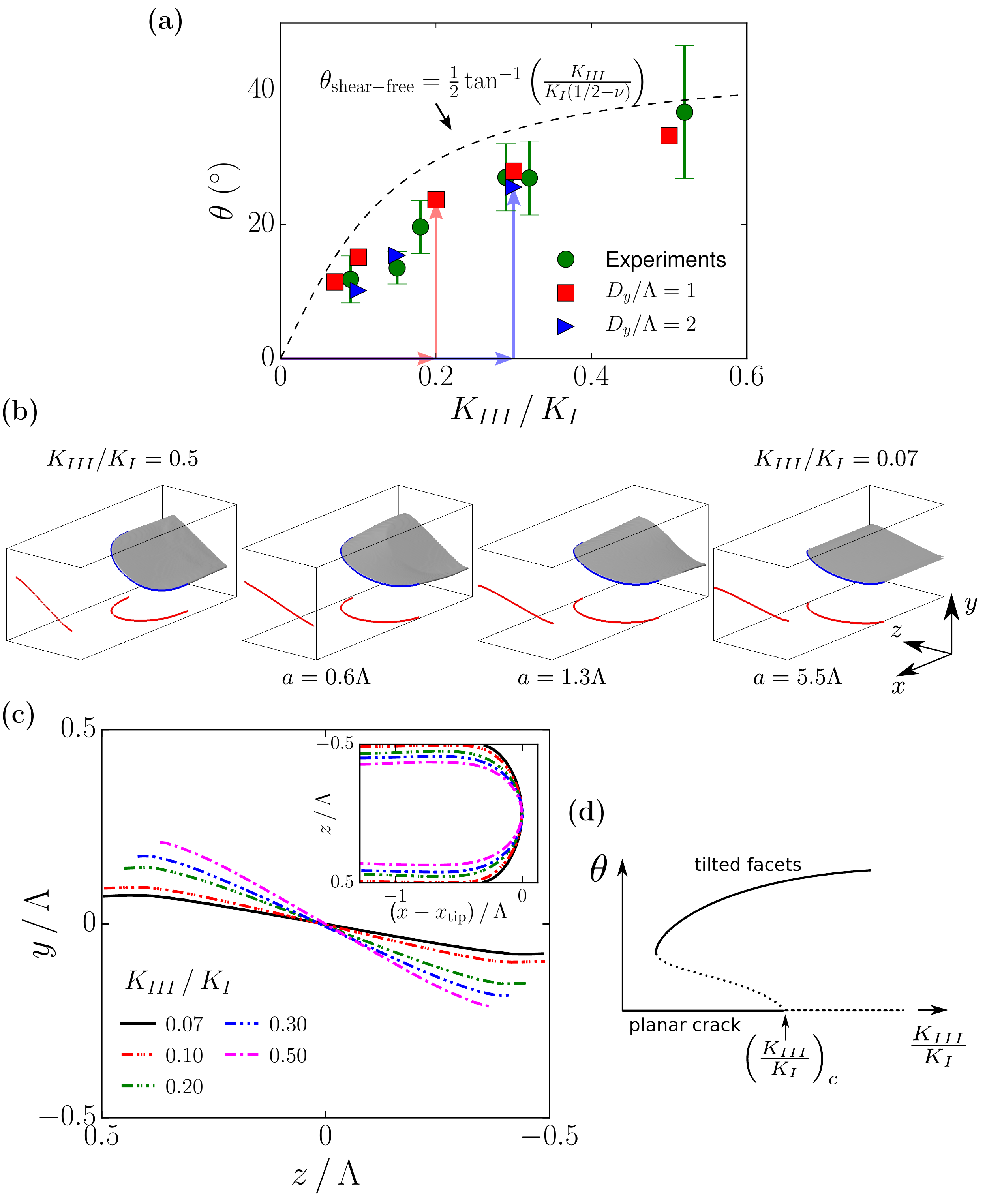}
\caption{ (Color online). (a) Comparison of facet tilt angles obtained from experiments and simulations, where red and blue arrows  indicate the instability thresholds of planar crack propagation for $D_y/\Lambda=1$ and $D_y/\Lambda=2$, respectively (see Fig. 2(e)), and  theoretically predicted assuming shear-free facets (dashed line) \cite{PK10,CP96}.
(b) Snapshots of a phase-field simulation for $D_y/\Lambda=1$ demonstrating the subcritical nature of the bifurcation from planar to segmented crack propagation.  A propagating segmented front solution for $K_{III}/K_{I}=0.5$ was used as initial condition ($\theta=31^\circ$). The facet continuously rotated towards a lower angle in response to the decrease in $K_{III}/K_{I}$ and then reached its steady state ($\theta=11.2^\circ$) after propagating a distance $a=5.5\Lambda$ (see Movie 2 of \cite{Suppinfo}). (c) Out-of-plane and in-plane  (inset) crack-front projections. In all simulations, $D_x = 154\xi$, $D_y= D_z=60\xi$, $\Lambda=60\xi$ and $G=1.5G_c$. (d) Schematic diagram of subcritical bifurcation recapitulating the experimental and simulations results with solid (dashed) lines representing stable (unstable) solutions.
}
\vskip -0.8cm
\label{Fig3}
\end{center}
\end{figure}

Next, in order to explore the nonlinear character of the bifurcation from planar to segmented crack front, we measured experimentally the facet tilt angle $\theta$ extracted from three-dimensional maps of post-mortem fracture surfaces obtained using a profilometer as detailed in \cite{CL2014}. The angle $\theta$ is plotted versus $K_{III}/K_I$ in Fig. \ref{Fig3}(a). Furthermore, 
we investigated computationally the propagation of periodic arrays of A facets, where coarsening is suppressed by choosing $D_z=\Lambda$ due to the periodic boundary conditions along $z$. In this geometry, we tracked the steady-state branch of propagating solutions by decreasing $K_{III}/K_I$ starting from values above the linear instability threshold to values below this threshold, as low as 0.07 to span the entire experimental range of mode mixity. For each $K_{III}/K_I$ value, we allowed the facet to relax to a new stationary shape and tilt angle, as illustrated in Fig. 3(b) for a simulation where $K_{III}/K_I$ was decreased from 0.5 to 0.07. The computed tilt angles are compared to experimental results in Fig. \ref{Fig3}(a) with the corresponding facet shapes shown in Fig. 3(c).
Both the facet shapes, which gently curve at their extremities in the $yz$ plane due to elastic interactions between neighboring facets, and the tilt angles are in good quantitative agreement with experimental observations within measurement errors. Fig. 3(a) also shows that 
computed tilt angles are weakly dependent on system size ($D_y/\Lambda$) and fall below the prediction of a simple theory, which assumes that facets are shear-free \cite{PK10,CP96}.
Those results demonstrate that propagating segmented front solutions exist over the entire range of $K_{III}/K_I$ investigated experimentally, including values less than $(K_{III}/K_I)_c$.
We conclude that the bifurcation from planar to segmented front is strongly subcritical, with bistability of planar and segmented crack growth for $K_{III}/K_I<(K_{III}/K_I)_c$ as illustrated schematically in Fig. 3(d).

\begin{figure}[t!]
\begin{center}
\includegraphics[width=0.47\textwidth]{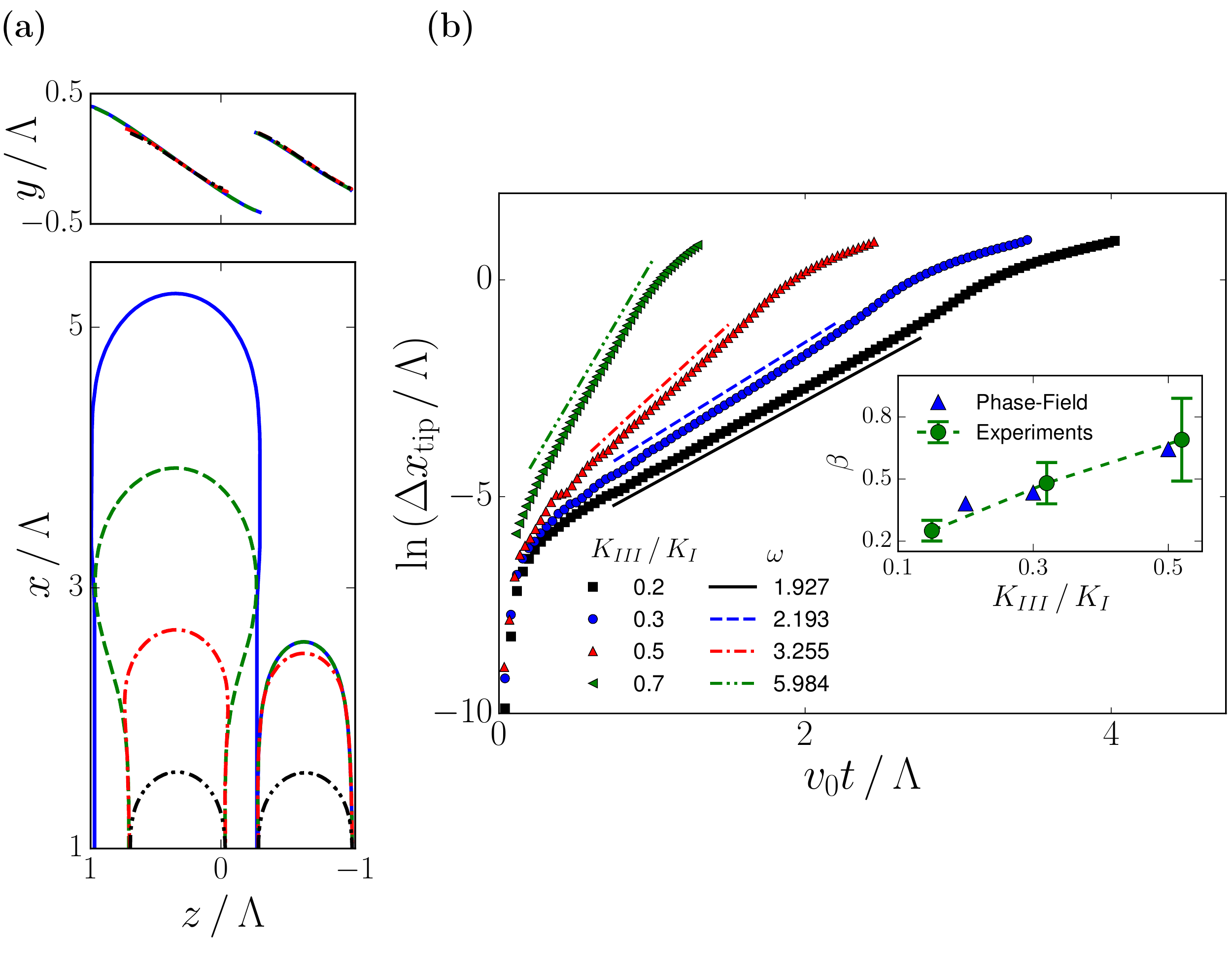} 
\caption{(Color online). (a) Illustration of spatial period doubling instability in a phase-field simulation for $K_{III}/K_I= 0.5$; out-of-plane and in-plane projections of crack fronts at different times are plotted in the top panel and the bottom panel, respectively (see Movie 3 of \cite{Suppinfo}).  (b) Semi-log plot of difference of tip positions along the propagation $x$-axis between leading and lagging facets versus scaled time for different $K_{III}/K_I$. Inset: coarsening rate $\beta$ versus $K_{III}/K_I$ obtained from experiments and phase-field simulations. In all simulations, $D_x = 307\xi$,  $D_y=60\xi$, $D_z  = 120 \xi$, $\Lambda=60\xi$ and $G=1.5G_c$.}
\vskip -0.8cm
\label{Fig-Coarsening}
\end{center}
\end{figure}

To characterize coarsening in phase-field simulations, we investigated the stability of periodic array of facets by repeating the above series of simulations with two facets ($D_z=2\Lambda$). This geometry is motivated by the striking similarity between the coarsening behavior of facets in the present experiments (Fig. 1(a)-(g)) and coarsening of curved fronts in other interfacial pattern forming systems, in particular viscous fingering \cite{KesLev1986} and dendritic crystal growth \cite{WarLan1993,Losertetal1996}. In those systems, it is well-established that coarsening of finger arrays is associated with a spatial period-doubling linear instability of the array leading to elimination of one of every two fingers in the array by exponential amplification of small perturbations. Results of simulations illustrated in Fig. \ref{Fig-Coarsening}(a) show that arrays of facets exhibit a similar spatial period doubling instability driven by elastic interactions between facets. This instability yields an increase (decrease) of the SIF and hence the energy release rate at the tips of leading (lagging) facets. The amplification rate of instability is obtained by computing the difference of $x$-tip position $\Delta x_{\rm tip}(t)$ between leading and lagging facets, which grows exponentially in time starting from an infinitesimal perturbation,  $\Delta x_{\rm tip}(t)\approx \Delta x_{\rm tip}(0)e^{\omega v_0t/\Lambda}$, where $v_0$ and 
$\Lambda$ are the initial facet growth velocity and spacing, respectively. The slopes of semi-log plots of $\Delta x_{\rm tip}(t)/\Lambda$ versus $v_0t/\Lambda$ in Fig. \ref{Fig-Coarsening}(b) yield values of $\omega$ that increase markedly with $K_{III}/K_I$, showing that a larger mode III component leads to a faster elimination rate of facets.

Coarsening, clearly visible in Fig. 1(b) and other experimental views \cite{Suppinfo}, was 
quantified experimentally by analyzing post-mortem fracture surfaces \cite{CL2014}. 
The results show that the relation between the mean facet spacing $\Lambda$ and the crack propagation length $a$ is approximately linear, with a mean slope $\beta\equiv d\Lambda/da$ increasing with $K_{III}/K_I$ (inset of Fig. \ref{Fig-Coarsening}(b)).
To relate the coarsening rates in phase-field simulations and experiments, we derive a simple evolution equation for the average array spacing $\Lambda$ based on dynamical mean-field picture as previously done for dendritic arrays \cite{WarLan1993}. The coarsening rate $\beta\equiv d\Lambda/da\approx \Delta\Lambda/\Delta a$ where $\Delta \Lambda$ is the change of array spacing due to elimination of one of every two facets along the array or $\Delta\Lambda\approx \Lambda$, while $\Delta a$ is the distance that the facets propagated during the elimination process. Since elimination occurs via exponential amplification of small perturbations, facets will propagate an average distance $\Delta a \sim \Lambda/\omega$ during this process, yielding the prediction $\beta \sim \omega$, or $\beta = C\omega$ where $C$ is a constant prefactor of order unity. The comparison in the inset of Fig. \ref{Fig-Coarsening}(b) shows that this simple theory is able to predict reasonably well the increase of the coarsening rate with $K_{III}/K_I$ up to the value of the constant prefactor $C = 0.198$ determined from a global best fit to the experimental data for all $K_{III}/K_I$ values.  

The present results reconcile the prediction of linear stability analysis (Eq. (2)) with experimental observations by showing that the bifurcation from planar to segmented crack growth is strongly subcritical; facet arrays exist as fundamental crack propagating solutions of LEFM for a range of $K_{III}/K_I$ values extending below the instability threshold. They further show that coarsening is driven by a spatial period doubling instability of facet arrays with a growth rate that depends on mode mixity. The reasonably good quantitative agreement between simulated and observed morphologies suggests that LEFM is an adequate theory to describe complex geometrical features of both brittle and fatigue cracks in mixed mode I+III fracture. While the present results show that the subcritical propagation of segmented cracks is theoretically possible, they do not identify the mechanism and scale of subcritical facet formation. As suggested by a recent LEFM analysis, 
materials imperfections may contribute to this process \cite{LL15}. However, this scenario, and even more fundamentally the ability of LEFM to model subcritical facet formation, remain to be explored both computationally and experimentally.


 \begin{acknowledgments}

The research at Northeastern University was supported by Grant No. DE-FG02-07ER46400 from the U.S. Department of Energy, Office of Basic Energy Sciences and a seed grant from the Massachusetts Green High Performance Computing Center. 
The research at University Paris Sud benefited of financial support from ANR GeoSMEC (2012-BS06-0016-03). The authors thanks L. Auffray,  D. Bonamy, F. Buchholz, V. Doquet, J.-C. Eytard, R. Pidoux, A. Tanguy for their help in the experiments and J.-B. Leblond for helpful discussions.
\end{acknowledgments}

\bibliography{MixedMode_bib}
 
\end{document}


\title{Supplemental Material for: Crack Front Segmentation and Facet Coarsening in Mixed-Mode Fracture}
\author{Chih-Hung Chen}
\affiliation{Physics Department and Center for Interdisciplinary Research on Complex 
Systems, Northeastern University, Boston, Massachusetts 02115, USA}
\author{Tristan Cambonie}
\affiliation{Laboratoire FAST, Univ Paris Sud, CNRS, Universit\'e Paris-Saclay, F-91405, Orsay, France}
\author{Veronique Lazarus}
\affiliation{Laboratoire FAST, Univ Paris Sud, CNRS, Universit\'e Paris-Saclay, F-91405, Orsay, France}
\author{Matteo Nicoli}
\affiliation{Physics Department and Center for Interdisciplinary Research on Complex 
Systems, Northeastern University, Boston, Massachusetts 02115, USA}
\author{Antonio Pons}
\affiliation{Department of Physics and Nuclear Engineering, Polytechnic University of Catalonia, Terrassa,
Barcelona 08222, Spain}
\author{Alain Karma}
\affiliation{Physics Department and Center for Interdisciplinary Research on Complex 
Systems, Northeastern University, Boston, Massachusetts 02115, USA}

\date{\today}

\maketitle

\section{Experiments}
\begin{figure}[h]
\begin{center}

{
\includegraphics[width=0.5\textwidth]{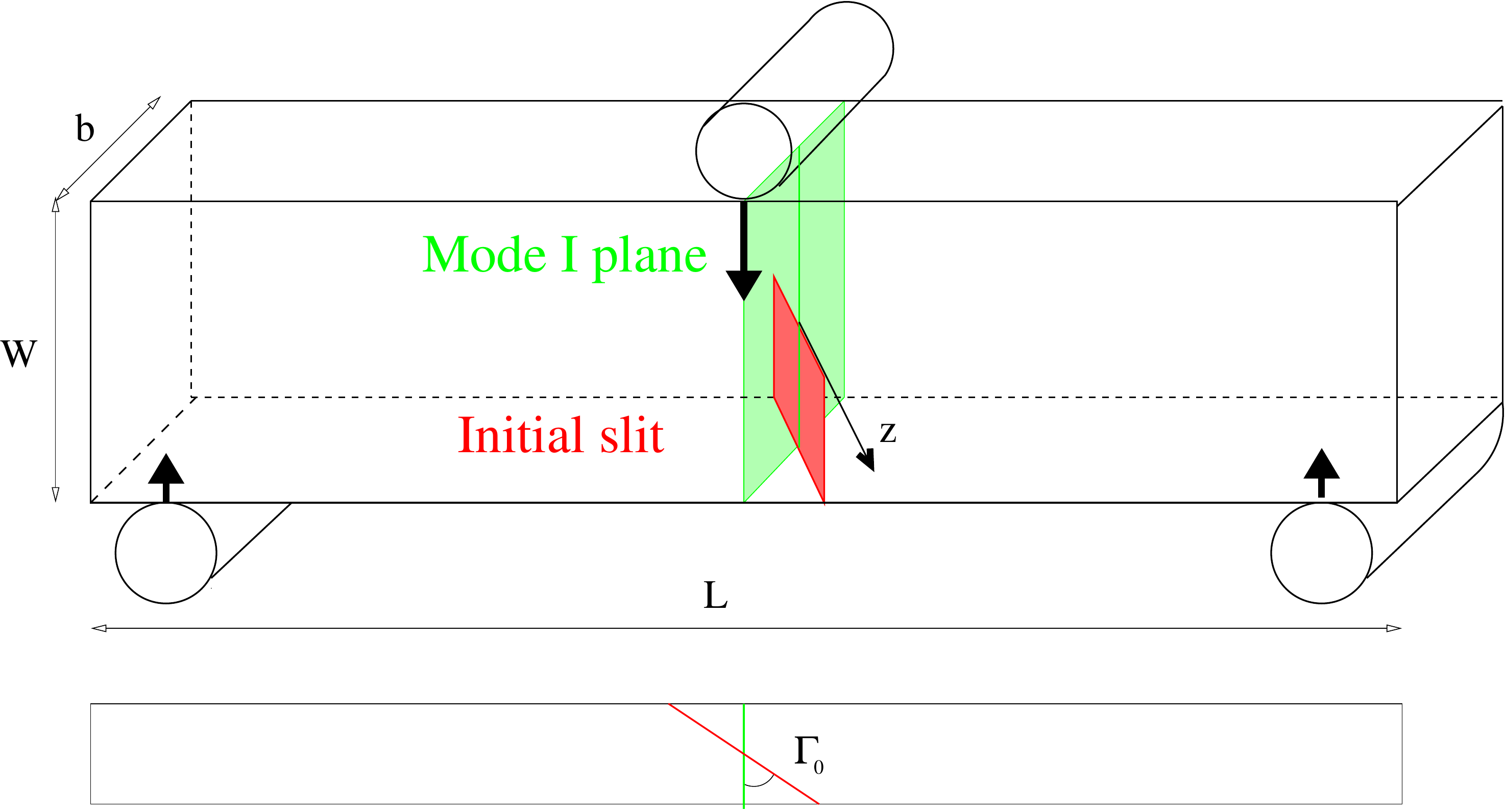} 
}
\hskip 15pt 
{
\includegraphics[width=0.1\textwidth]{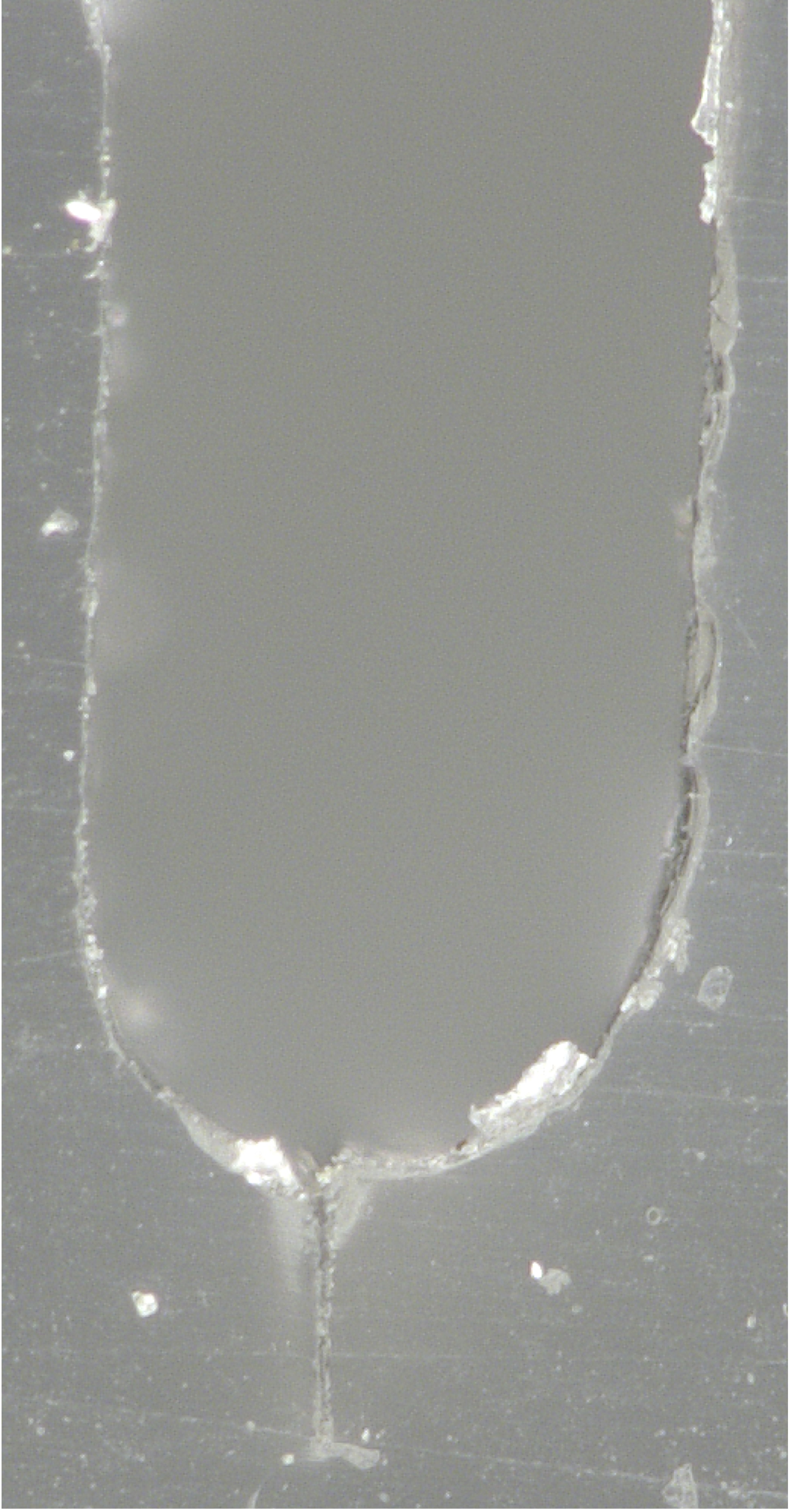}  
}

\caption{Left hand side: 3 Point Bending setup. Right hand side: Lateral view of the initial slit made by micro-milling and of the additional sharp crack introduced by pushing under a controlled manner, a razor blade in the initial slit.}
\label{fig:setup}
\end{center}
\end{figure}

Experiments were carried out using plexiglas beams of dimensions $L\times W\times b$  (Fig.  \ref{fig:setup}, left). 
We used two different sizes corresponding to $L=260$ mm, $W=60$ mm, $b=10$ mm for the large ones and $L=100$ mm, $W=10$ mm, $b=10$ mm for the small ones. The initial slit with a blunted tip of radius $R=300 \: \mu$m was made by micro-milling.   To initiate a  true crack  with a sharp tip and a smooth straight front (Fig. \ref{fig:setup}, right), we pushed a wedge (razor blade) into the slit, quasi-statically, under controlled slowly increasing force, using a tensile machine. The total length $d$ of the crack (slit+sharp crack)  is  $d\sim W/3$ in both cases. The  residual stresses introduced by the slit manufacturing, were relaxed by heating the samples at 90-95 $^\circ$C during 10 hours. Their annihilation is checked using polarizers.  To introduce some amount of mode III, the initial planar notch in the sample is tilted at an angle $\Gamma_{0}$ from the mode I central plane of symmetry, $\Gamma_{0}$ varying between $15^\circ$ and $45^\circ$, where zero angle corresponds to pure mode I loading.  Larger and smaller samples were loaded in 3 and 4 point bending setups, respectively. 
 The loading frequency was $f=5$ Hz and $K_{min}/K_{max} = 0.1$. The amplitude of the stress intensity factor $\Delta K \sim 0.5$ MPa$\cdot$m$^{1/2}$ 
 has been chosen well below the brittle fracture threshold $K_{c} \sim 1$ MPa$\cdot$m$^{1/2}$, to avoid  brittle fracture, while being large enough to ensure propagation \cite{PulKna98}.

\begin{figure}
\begin{center}
\hskip -5pt  
\includegraphics[width=0.98\textwidth]{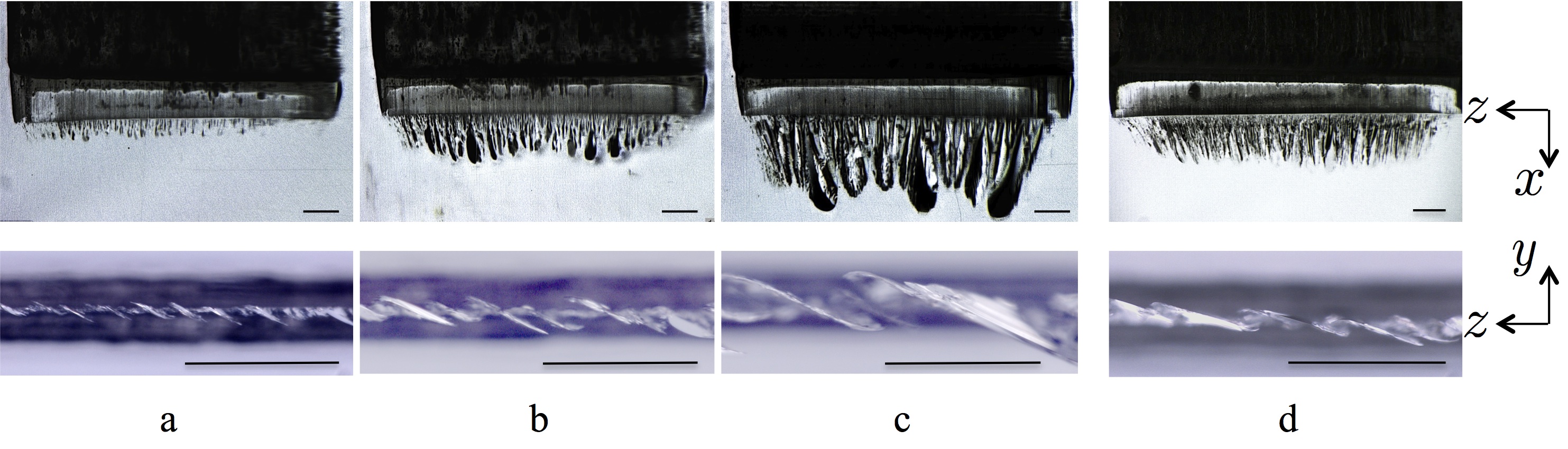} 
\caption{In-situ numerical microscope images of partially broken samples. Each column corresponds to one sample, the first row being a front view and the second row being a bottom view as indicated by cartesians axes corresponding to Fig. 1(h) of the main text. 
  $K_{III}/K_I\approx 0.3$ in (a)-(c) and $\approx 0.18$ in (d). In the front views, the initial slit and the facets appear in black. In the bottom views, the initial larger slit is dark and the facets appear in white. The bar scale is 1 mm in all images.
}
\label{fig:insitu}
\end{center}
\end{figure}

Sets of experiments were performed on  series of similar beams by stopping the propagation at different stages of the propagation. Figure \ref{fig:insitu} gives some in-situ views of the corresponding crack morphologies, acquired with a numerical Keyence microscope. Front and bottom views are given in the first and second row respectively. Each column corresponds to the same sample. The beam size is $L=100$ mm, $W=10$ mm, $b=10$ mm; columns (a)-(c) correspond to $\Gamma_{0}=30^\circ$ ($K_{III}/K_I\approx 0.3$) and column (d) to $\Gamma_{0}=20^\circ$  ($K_{III}/K_I\approx 0.18$).
The development of rotated facets that coalesce during propagation are clearly visible.  One can observe i) comparing columns (a) to (c), that the rotation angle of those facets is approximately constant during propagation, for a given value of $K_{III}/K_{I}$ and  ii) comparing columns (a)-(c) with (d), that the rotation angle of these facets decreases with decreasing $K_{III}/K_{I}$.

\begin{figure}[t!]
\begin{center}
\hskip -5pt  
\includegraphics[width=\textwidth]{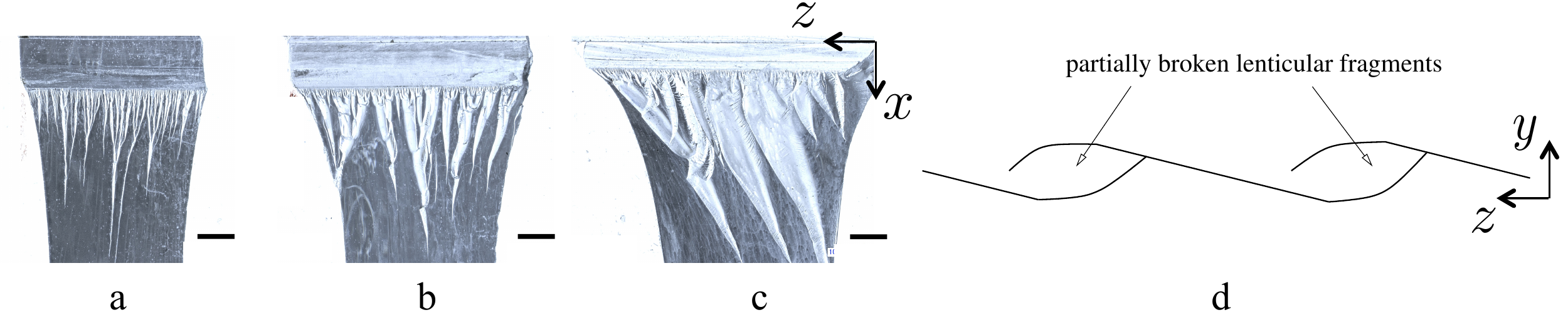}
\caption{Post-mortem front views of the fracture surfaces corresponding to (a) $K_{III}/K_{I}=0.15$,  (b) $K_{III}/K_{I}=0.3$,  (c) $K_{III}/K_{I}=0.5$ ($L=260$ mm, $W=60$ mm, $b=10$ mm). The white zones corresponds to the partially broken fragments left behind on the crack surface. The formation of these fragments is due to the lateral propagation of two adjacent interacting facets as sketched on (d). The bar scale is 2 mm in (a)-(c).}
\vskip -20pt 
\label{fig:postmortem}
\end{center}
\end{figure}

The coalescence rate $\beta$ cannot be quantified on the in-situ samples due to optical distortions induced by the plexiglas.
For this purpose, other samples were broken completely (Fig. \ref{fig:postmortem} (a)-(c)) and three dimensional profilometer maps of their fracture surfaces were done. Quantification of $\theta$ (Fig. 3(a) of the main text) and $\beta$ (inset of Fig. 4(b) of the main text) has been done performing some statistical post-treatment over several facets of these profiles.  The procedure is explained in detail in \cite{CamLaz14ECF20p}.

The more clear zones in Fig. \ref{fig:postmortem}(a)-(c) correspond to fragments left behind by partial breaking of the zones located between adjacent facets.
Indeed, beside propagating in the $x-$direction,  the facets propagate also in the lateral $z-$direction. 
As they interact, they curve to form the well-known \cite{Mel83, DanFenLec10} ``en-passant S-shape'' pattern (see Fig. \ref{fig:postmortem}(d)).
Since the symmetry of this interaction is sensible to any perturbation, generally only one out of two crack tips connects to the adjacent facet, leaving on one
of the fracture surface,
a partially broken fragment. These fragments are visible in white on Fig. \ref{fig:postmortem}(a)-(c) and are lacking on the complementary fracture surface.

\section{Simulations}
We conducted large-scale simulations of mixed-mode I+III fracture using the phase-field model originally proposed by Karma, Kessler and Levine (KKL) \cite{KKL01}, which has been used to solve fracture propagation problems over the last decade \cite{KL04,HL04,HA13,H08,PK10}. The model introduces a scalar order parameter $\phi$ to distinguish between intact and broken states of the material. The total energy of the system is given by the functional \cite{KKL01,HK09} 

\begin{equation}
E = \int\left\{ \frac{\rho}{2}\left(\frac{\partial\vec{u}}{\partial t}\right)^{2}+\frac{\kappa}{2}\left(\nabla\phi\right)^{2}+g\left(\phi\right)\left(e_{{\rm strain}}-e_{{\rm c}}\right)\right\} dV,\label{eq:EnergyFunctional}
\end{equation}
where $\rho$ is the mass density; $g\left(\phi\right)=4\phi^{3}-3\phi^{4}$ is conventionally chosen with the properties that $g\left(0\right)=0$, $g\left(1\right)=1$ and $g'\left(1\right)=g'\left(0\right)=0$ \cite{KKL01,HK09}; $\vec{u}=\left(u_{x},u_{y},u_{z}\right)$ represents the displacement field; 
 $e_{\rm{strain}}=\lambda\left(\varepsilon_{ii}\right)^{2}/2+\mu\left(\varepsilon_{ij}\right)^{2}$  is the strain energy density and $\varepsilon_{ij}=\left(\partial_{i}u_{j}+\partial_{j}u_{i}\right)/2$ is the strain tensor of a linear elastic material with $i=1$, 2 and 3 corresponding to $x$, $y$ and $z$, respectively, $\lambda$ and $\mu$ are the Lam\'e coefficients. The function $g$ decreases monotonously from the intact state ($\phi=1$) to the fully broken state ($\phi=0$) and accounts for elastic softening at large strain. When the strain energy density $e_{\rm{strain}}$ exceeds the threshold $e_{c}$, the broken state becomes energetically favored. Energy dissipation occurs in the process zone of size $\xi=\sqrt{\kappa/\left(2e_{c}\right)}$ around the crack tip, where $\phi$ varies smoothly
between 0 and 1. In the phase-field model, the Griffith's criterion $G_{c}$ is given by $G_{c}=2\sqrt{2\kappa e_{c}}\int_{0}^{1}\phi\sqrt{1-g\left(\phi\right)}$ \cite{KKL01,HK09}; the iso-surfaces of $\phi=0.5$ are conventionally defined as the fracture surfaces.
The equations of motion for the phase-field $\phi$ and displacement field $\vec{u}$ are derived variationally from the energy functional \cite{KKL01,HK09}: 

\begin{equation}
\frac{\partial\phi}{\partial t}=-\chi\frac{\delta E}{\delta\phi}, \label{eq:eom-phi}
\end{equation}
\begin{equation}
\rho\frac{\partial^{2}u_{i}}{\partial t^{2}}=-\frac{\delta E}{\delta u_{i}}. \label{eq:eom-u}
\end{equation}

We studied crack-front instabilities of mixed mode I+III fracture propagation along the $x$-axis in a rectangular slab of length $D_x$, width $D_y$ and thickness $D_z$ with center at the origin.  The mixed I+III loading was imposed by fixed displacements at $y=\pm D_y/2$, $u_{y}\left(x,\pm D_y/2,z \right)=\pm\Delta_{y}$ (mode I) and $u_{z}\left(x,\pm D_y/2,z\right)=\pm\Delta_{z}$ (mode III), and periodic boundary conditions in the $z$-direction. To allow full relaxation of the crack front to a stationary state, simulations were carried out on a ``treadmill'' that adds a strained $(y,z)$ layer at $x=D_x/2$ and removes a layer at $x=-D_x/2$ when the crack has advanced by one lattice spacing. 
This allows us to simulate crack propagation lengths much longer than $D_x$ ($a\gg D_x$), thereby modeling propagation in a slab infinitely long in $x$. 

\begin{figure}
\includegraphics[width=0.4\paperwidth]{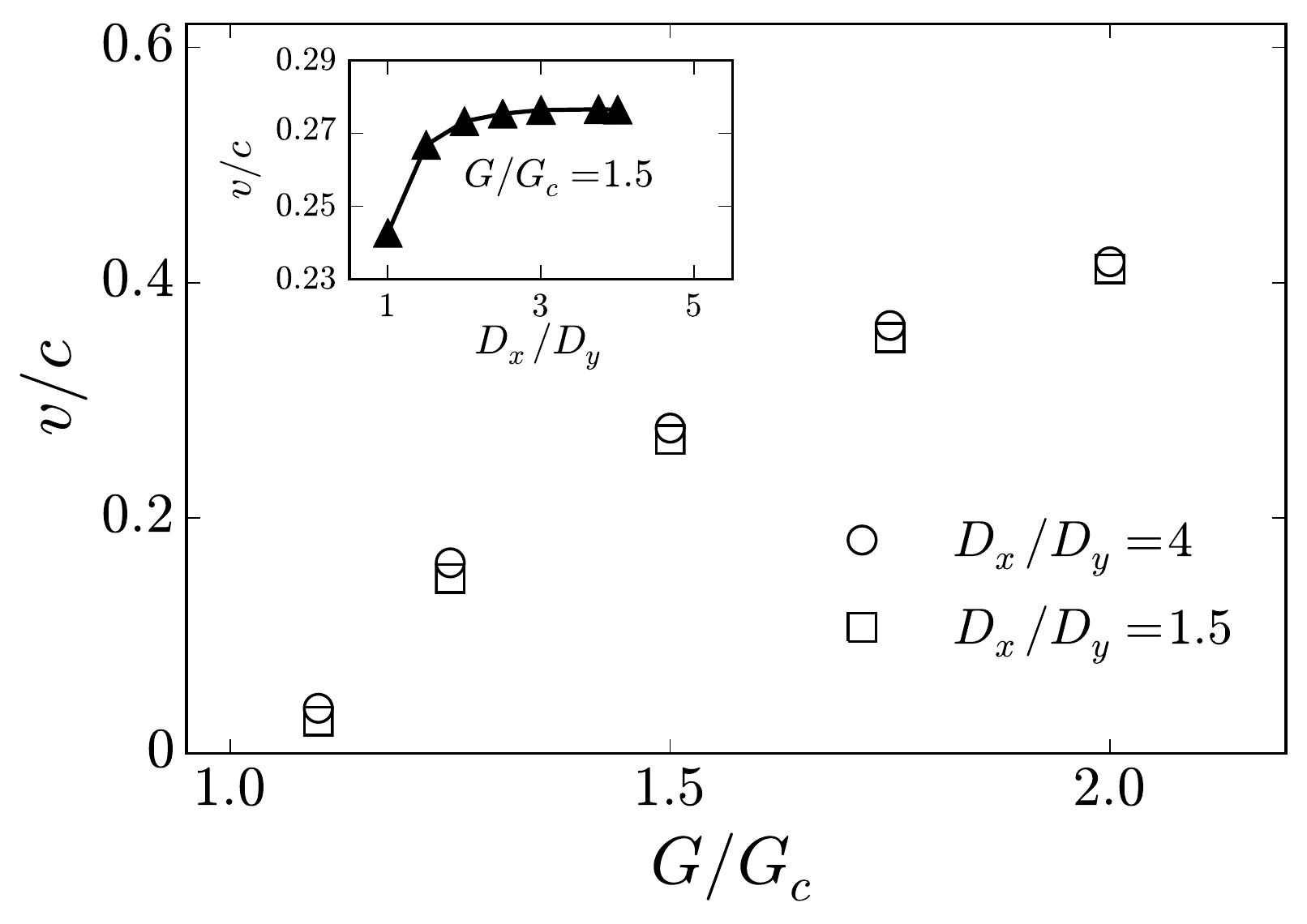}
\caption{Plot of pure mode I crack propagation speed $v$ (scaled by the shear wave speed $c$) versus $G/G_{c}$. Phase-field simulations show that the crack speed becomes independent of the simulation box length when $D_x/D_y\ge2.5$ (inset). Simulation parameters are $D_y=60\xi$ and $D_z=30\xi$.
\label{fig:Crack-Speeds} }
\end{figure}

The crack dynamics is controlled by two key parameters $K_{III}/K_{I}$ and $G/G_{c}$, where $K_{I}=2\mu\Delta_{y}\sqrt{2 \left(1+\lambda/\mu\right)/D_y}$ and $K_{III}=2\mu\Delta_{z}/\sqrt{D_y}$ are the stress intensity factors of a semi-infinite planar crack and $G=\left(\left(1-\nu\right)K_{I}^{2}+K_{III}^{2}\right)/\left(2\mu\right)$ is the corresponding energy release rate, where $\nu$ is Poisson's ratio.
Simulations were carried out with the range $G/G_{c}\le1.5$ where the ratio of the crack propagation speed to the shear wave speed $v/c\le0.3$, which is small enough to neglect inertial effects. 
The slab length $D_x$ was chosen greater than $2.5$ of the slab width $D_y$ to eliminate the influence of the two end-boundaries of the slab ($x=\pm D_x/2$) on the central region of the slab ($|x| \ll D_x$) where the average crack front position is maintained by the treadmill. 
As a result, the planar crack propagation speed becomes independent on the system length, as shown in Fig. \ref{fig:Crack-Speeds}. 

We rewrite the phase-field model in a dimensionless form by measuring length in units of the fracture process zone scale $\xi$ and time in units of the characteristic dissipation time scale $\tau=1/\left(\mu\chi\right)$. 
The remaining simulation parameters are chosen as follows, $e_{c}/\mu=1/2$, $\nu=0.38$ and $c\tau/\xi=2.76$, where the shear wave speed $c=\sqrt{\mu/\rho}$. Large scale simulations of the order of $10^{7}-10^{8}$ grid points were performed using graphics processing units (GPUs) with the CUDA parallel programming language.

\begin{figure}
\includegraphics[width=0.3\paperwidth]{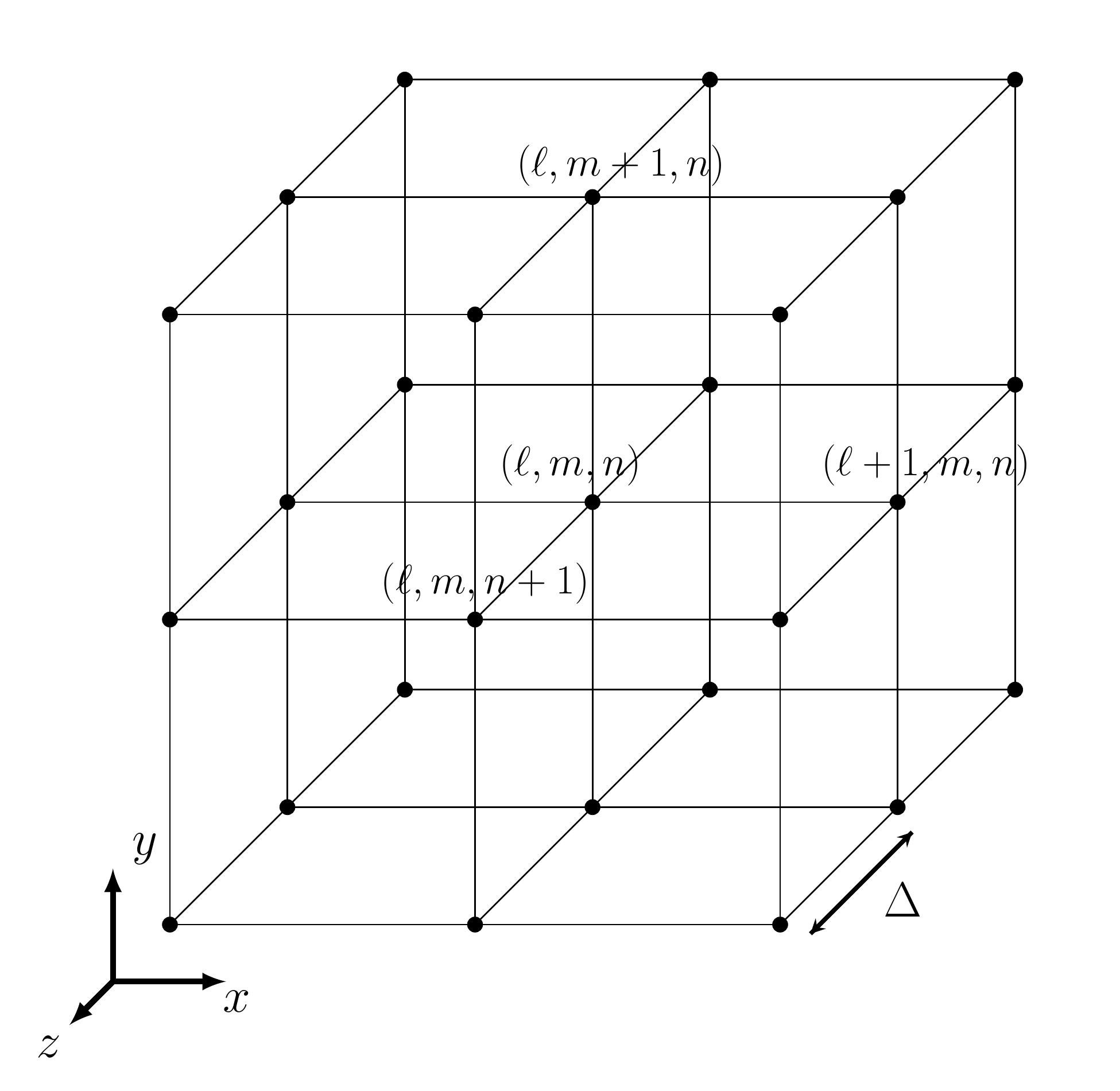}
\caption{
Spatial discretization of phase-field model on a cubic grid of uniform mesh size $\Delta$. Gridpoint labeled $(\ell, m, n)$ represents a unit cubic of material of mass $\rho\Delta^3$ at location $(\ell\Delta, m\Delta, n\Delta)$. }
\label{fig:Cubic-grid}
\end{figure}

The energy functional of the phase-field model was discretized in space on a cubic grid of uniform mesh size $\Delta=0.3\xi$ (Fig. \ref{fig:Cubic-grid}). To write the discretized energy in compact form, we define the superscripts $\ell$, $m$, and $n$ to refer to the gridpoint at $x=\ell\Delta$, $y=m\Delta$, and $z=n\Delta$, the subscripts $\left\{ i,j\right\} =\left\{1,2,3\right\}$ to refer to the corresponding $\left\{ x,y,z\right\} $ components of the displacement fields, and use the standard Kronecker delta defined by $\delta_{ij}=1$ if $i=j$ and $\delta_{ij}=0$ if $i\ne j$. 
We discretize ${\partial u_{i}} /{\partial x_{j}}$, $\left({\partial u_{i}} / {\partial x_{j}}\right)^2$ and $\left({\partial \phi}/{\partial x_{j}}\right)^2$ on the gridpoint $\left(\ell,m,n\right)$ by the following approximations:

\begin{eqnarray*}
\left(\frac{\partial u_{i}}{\partial x_{j}}\right)\left(\ell,m,n\right) & \approx & \frac{1}{2\Delta}\left(u_{i}^{\ell+\delta_{1j},m+\delta_{2j},n+\delta_{3j}}-u_{i}^{\ell-\delta_{1j},m-\delta_{2j},n-\delta_{3j}}\right),
\end{eqnarray*}

\begin{eqnarray*}
\left(\frac{\partial u_{i}}{\partial x_{j}}\right)^{2}\left(\ell,m,n\right) & \approx & \frac{1}{2\Delta^{2}}\left[\left(u_{i}^{\ell+\delta_{1j},m+\delta_{2j},n+\delta_{3j}}-u_{i}^{\ell ,m,n}\right)^{2}+\left(u_{i}^{\ell ,m,n}-u_{i}^{\ell-\delta_{1j},m-\delta_{2j},n-\delta_{3j}}\right)^{2}\right],
\end{eqnarray*}
and 
\begin{eqnarray*}
\left(\frac{\partial \phi}{\partial x_{j}}\right)^{2}\left(\ell,m,n\right) & \approx & \frac{1}{2\Delta^{2}}\left[\left(\phi^{\ell+\delta_{1j},m+\delta_{2j},n+\delta_{3j}}-\phi^{\ell ,m,n}\right)^{2}+\left(\phi^{\ell ,m,n}-\phi^{\ell-\delta_{1j},m-\delta_{2j},n-\delta_{3j}}\right)^{2}\right].
\end{eqnarray*}

Accordingly, the strain energy density $e_{\rm{strain}}$ on gridpoint $\left(\ell,m,n\right)$ becomes

\begin{eqnarray*}
e_{\rm{strain}}^{\ell, m,n} & = & \frac{1}{4\Delta^{2}} \Big(\left(\lambda+2\mu\right)\left(H_{1}^{\ell ,m,n}+H_{2}^{\ell ,m,n}+H_{3}^{\ell ,m,n}\right)+\lambda\left(I_{1,2}^{\ell ,m,n}+I_{2,3}^{\ell ,m,n}+I_{3,1}^{\ell ,m,n}\right)\\
 &  & +\mu \left(J_{1,2}^{\ell ,m,n}+J_{2,3}^{\ell ,m,n}+J_{3,1}^{\ell ,m,n}\right) \Big)  .
\end{eqnarray*}

where we have defined the functions

\begin{equation*}
H_{i}^{\ell ,m,n}\equiv\left(u_{i}^{\ell+\delta_{1i},m+\delta_{2i},n+\delta_{3i}}-u_{i}^{\ell ,m,n}\right)^{2}+\left(u_{i}^{\ell ,m,n}-u_{i}^{\ell-\delta_{1i},m-\delta_{2i},n-\delta_{3i}}\right)^{2},
\end{equation*}

\begin{equation*}
I_{i,j}^{\ell ,m,n}\equiv\big(u_{i}^{\ell+\delta_{1i},m+\delta_{2i},n+\delta_{3i}}-u_{i}^{\ell-\delta_{1i},m-\delta_{2i},n-\delta_{3i}}\big)\big(u_{j}^{\ell+\delta_{1j},m+\delta_{2j},n+\delta_{3j}}-u_{j}^{\ell-\delta_{1j},m-\delta_{2j},n-\delta_{3j}}\big),
\end{equation*}
and

\begin{eqnarray*}
J_{i,j}^{\ell ,m,n} & \equiv & \left(u_{i}^{\ell+\delta_{1j},m+\delta_{2j},n+\delta_{3j}}-u_{i}^{\ell ,m,n}\right)^{2}+\left(u_{i}^{\ell ,m,n}-u_{i}^{\ell-\delta_{1j},m-\delta_{2j},n-\delta_{3j}}\right)^{2}\\
 &  & +\left(u_{j}^{\ell+\delta_{1i},m+\delta_{2i},n+\delta_{3i}}-u_{j}^{\ell ,m,n}\right)^{2}+\left(u_{j}^{\ell ,m,n}-u_{j}^{\ell-\delta_{1i},m-\delta_{2i},n-\delta_{3i}}\right)^{2}\\
 &  & +\left(u_{i}^{\ell+\delta_{1j},m+\delta_{2j},n+\delta_{3j}}-u_{i}^{\ell-\delta_{1j},m-\delta_{2j},n-\delta_{3j}}\right)\left(u_{j}^{\ell+\delta_{1i},m+\delta_{2i},n+\delta_{3i}}-u_{j}^{\ell-\delta_{1i},m-\delta_{2i},n-\delta_{3i}}\right).
\end{eqnarray*}

The discretized form of the energy functional Eq. \eqref{eq:EnergyFunctional} on the cubic grid becomes 

\begin{eqnarray}
\frac{E}{\Delta^{3}}\approx{\displaystyle \sum_{\ell,m,n}} \left(\sum_{i= 1} ^{3}\frac{\rho}{2}\left(\partial_{t}u_{i}^{\ell ,m,n}\right)^{2}+\sum_{i= 1}^{3}\frac{\kappa}{4\Delta^{2}}F_{i}^{\ell ,m,n} + g\left(\phi^{\ell ,m,n}\right)\left(e_{\rm{strain}}^{\ell ,m,n}-e_{c}\right)\right), \label{eq:Energy-discrete}
\end{eqnarray}
where, in addition to the functions defined above,
\begin{equation*}
F_{i}^{\ell ,m,n}\equiv\left(\phi^{\ell+\delta_{1i},m+\delta_{2i},n+\delta_{3i}}-\phi^{\ell ,m,n}\right)^{2}+\left(\phi^{\ell ,m,n}-\phi^{\ell-\delta_{1i},m-\delta_{2i},n-\delta_{3i}}\right)^{2}.
\end{equation*}

From Eq. \eqref{eq:eom-phi}, Eq. \eqref{eq:eom-u} and Eq. \eqref{eq:Energy-discrete},  we obtain the equations of motion 
\begin{equation}
\frac{\partial\phi^{\ell ,m,n}}{\partial t}=-\chi\frac{\partial \left( E/ \Delta^{3} \right) }{\partial\phi^{\ell ,m,n}},\label{eq:eom-phi-discrete}
\end{equation}
\begin{equation}
\rho\frac{\partial^{2}u^{\ell ,m,n}_{i}}{\partial t^{2}}=-\frac{\partial  \left(E/ \Delta^{3} \right)}{\partial u^{\ell ,m,n}_{i}},\label{eq:eom-u-discrete}
\end{equation}
which are integrated in time using a Verlet scheme with a timestep size $\Delta t/\tau=0.001$. 

\begin{figure}
\includegraphics[width=0.35\paperwidth]{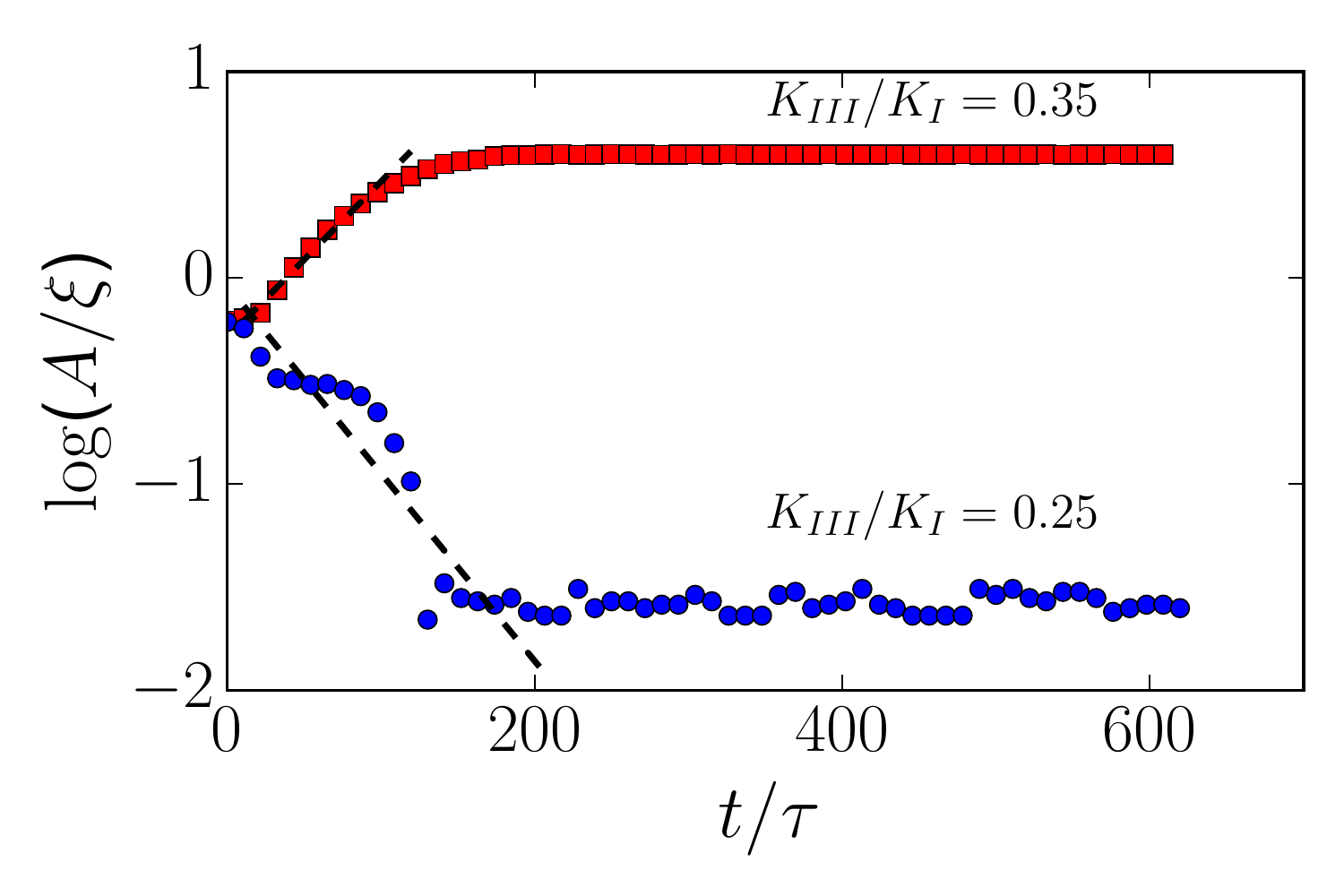}
\caption{(Color online). Semi-log plots of perturbation amplitude $A$ versus scaled time in simulations. The perturbation amplitude $A$ exponentially grew in time for $K_{III}/K_{1}=0.35$ (red squares) and decayed in time for $K_{III}/K_{1}=0.25$ (blue circles). Simulation parameters are $G=1.5G_c$, $A_0=0.6\xi$, $D_x=230\xi$, $D_y=120\xi$ and $D_z=30\xi$.  The same linear instability threshold was obtained in simulations using $A_0=1.2\xi$ and $A_0=1.8\xi$.
\label{fig:Linear-Instability} }
\end{figure}

To investigate the onset of instability of mixed mode I+III crack propagation, we carried out simulations starting from a parent planar crack in $yz$ plane for $x\le 0$. As in \cite{PK10}, the planar crack front was initially perturbed by a helical perturbation, $\delta x_{\rm{front }}+i\delta y_{\rm{front }}=A_{0}e^{-ikz}$, where $\delta x_{\rm{front }}$ and $\delta y_{\rm{front }}$ indicate the $x$ and $y$ components of deviations of the front from the reference planar crack, respectively. The amplitude $A\left(t\right)=A_{0}e^{\sigma t/\tau}$ was then found to grow $\left(\sigma>0\right)$  or decay $\left(\sigma<0\right)$ exponentially in time (Fig. \ref{fig:Linear-Instability}).

\begin{figure}
\includegraphics[width=0.7\paperwidth]{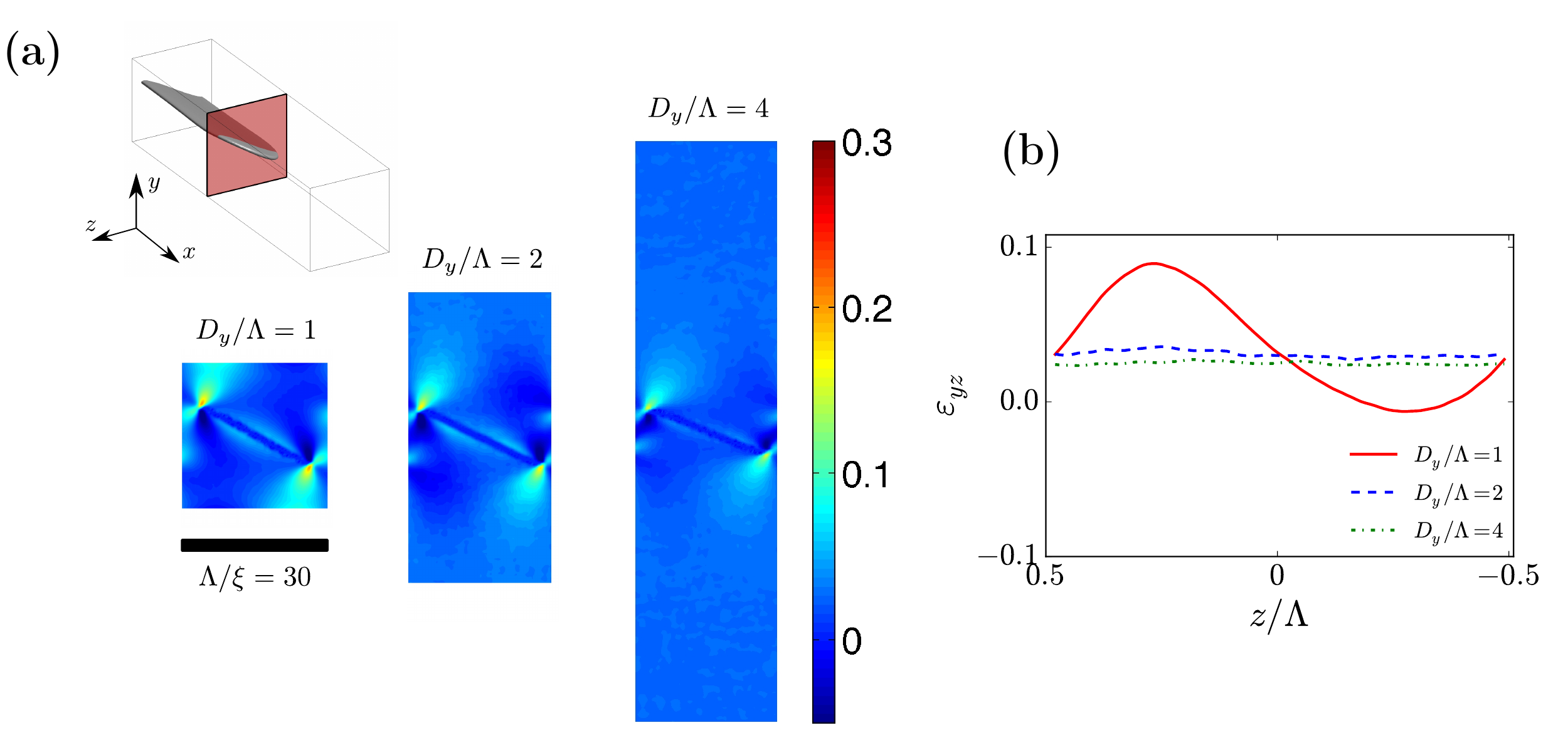}
\caption{Finite size effects: (a) (Color online). Images of strain field $\varepsilon_{yz}$ of a slice along $x$ behind the facet tips (the corresponding slice position is illustrated in the top panel) from simulations with $D_y/\Lambda=1$, 2 and 4. (b) Plots of strain field $\varepsilon_{yz}$ of the sliced regions at $y=D_y/2$ versus $z$ (scaled by $\Lambda$). The strain fields on boundaries are strongly influenced by facets when $D_y/\Lambda=1$  (red line) and nearly constant along $z$ for both $D_y/\Lambda=2$ (blue dashed line) and  $D_y/\Lambda=4$ (green double dashed line), where boundary effects become negligible.
In all simulations, $K_{III}/K_{I}=0.35$, $D_x/D_y=3.75$ and $D_z=\Lambda=30\xi$.
} 
\label{fig:Strain-Fields} 
\end{figure}

The theoretical instability threshold predicted by Eq. (2) of the main text is only valid in the large system size limit \cite{LKL11}. To characterize finite size effects, simulations with $D_y/\Lambda$ between 0.5 and 4 were carried out. The results show that finite size effects reduce significantly $\left(K_{III}/K_{I}\right)_c$ when $D_y$ is comparable to $\Lambda$ (Fig. 2(e) of the main text). For such system sizes, the strain fields along the $y=\pm D_y/2$ boundaries are spatially varying along $z$ and hence strongly influenced by the facets. However for $D_y/\Lambda \ge 2$, the strain fields become nearly independent of $z$ as expected in the large system size limit $D_y\gg \Lambda$ (Fig. \ref{fig:Strain-Fields}). In addition, the results of simulations with $\Lambda/\xi$ between 30 and 60 show that the crack-front instability is independent of $\Lambda$ in the limit $\Lambda\gg \xi$. 

\section{Acknowledgments}
\begin{acknowledgments}
The research at Northeastern University was supported by Grant No. DE-FG02-07ER46400 from the U.S. Department of Energy, Office of Basic Energy Sciences and a seed grant from the Massachusetts Green High Performance Computing Center. 
The research at University Paris Sud benefited of financial support from ANR GeoSMEC (2012-BS06-0016-03). 
We thank posthumously the late F. Buchholz, who has broken the largest samples, we have further exploited here. We also thank L. Auffray,  D. Bonamy, V. Doquet, J.-C. Eytard, R. Pidoux, A. Tanguy  for their help in the experiments.
\end{acknowledgments}

\bibliography{MixedMode_bib}